\documentclass[aps,prb,twocolumn,superscriptaddress]{revtex4}
\usepackage{graphicx}
\usepackage{amssymb}
\usepackage{amsmath}
\usepackage{graphicx}
\usepackage{epsfig}
\usepackage{color}
\usepackage{bm}
\usepackage{booktabs}
\usepackage{array}

\begin{document}

\title{Nonlocal Kondo effect and two-fluid picture revealed in an exactly solvable model}
\author{Jiangfan Wang}
\affiliation{Beijing National Laboratory for Condensed Matter Physics, Institute of Physics,
Chinese Academy of Sciences, Beijing 100190, China}
\author{Yi-feng Yang}
\email[]{yifeng@iphy.ac.cn}
\affiliation{Beijing National Laboratory for Condensed Matter Physics,  Institute of Physics, 
Chinese Academy of Sciences, Beijing 100190, China}
\affiliation{School of Physical Sciences, University of Chinese Academy of Sciences, Beijing 100190, China}
\affiliation{Songshan Lake Materials Laboratory, Dongguan, Guangdong 523808, China}
\date{\today}

\begin{abstract}
Understanding the nature of local-itinerant transition of strongly correlated electrons is one of the central problems in condensed matter physics. Heavy fermion systems describe the $f$-electron delocalization through Kondo interactions with conduction electrons. Tremendous efforts have been devoted to the so-called Kondo-destruction scenario, which predicts a dramatic local-to-itinerant quantum phase transition of $f$-electrons at zero temperature. On the other hand,  two-fluid behaviors have been observed in many materials, suggesting coexistence of local and itinerant  $f$-electrons over a broad temperature range but lacking a microscopic theoretical description. To elucidate this fundamental issue, here we propose an exactly solvable Kondo-Heisenberg model in which the spins are defined in the momentum space and the $\textbf{k}$-space Kondo interaction corresponds to a highly nonlocal spin scattering in the coordinate space. Its solution reveals a continuous evolution of the Fermi surfaces with Kondo interaction and two-fluid behaviors similar to those observed in real materials. The electron density violates the usual Luttinger's theorem, but follows a generalized one allowing for partially enlarged Fermi surfaces due to partial Kondo screening in the momentum space. Our results highlight the consequence of nonlocal Kondo interaction relevant for strong quantum fluctuation regions, and provide important insight into the microscopic description of two-fluid phenomenology in heavy fermion systems.
\end{abstract}

\maketitle

\section{Introduction}

Underlying the rich emergent quantum phenomena of heavy fermion systems \cite{Stockert2011, Pfleiderer2009} is the local-to-itinerant transition of $f$-electrons controlled by the interplay of Kondo and Ruderman-Kittel-Kasuya-Yosida (RKKY) interactions \cite{YangReview2022, Si2001, Coleman2001, Senthil2004,Pepin2005,PaulPepin2007,Komijani2019, Wang2020_local, Wang2021_nonlocal, Wang2022_Z2, Wang2022_FM, Wang2022_fermion, DongPRB2022}. Below the so-called coherence temperature $T^*$, a large amount of experimental observations have pointed to the coexistence of local and itinerant characters of $f$-electrons as captured phenomenologically by the two-fluid model \cite{Nakatsuji2004, Curro2004,YangPRL2008, YangNature2008, YangPNAS2012, Curro2012PNAS, YangReview2016}, which assumes the coexistence  of an itinerant heavy electron fluid formed by hybridized (screened) $f$-moments and a (classical) spin liquid of residual unhybridized $f$-moments. The two-fluid behavior exists over a broad temperature range, from the normal state below the coherence temperature down to inside the quantum critical superconducting phase \cite{YangPRL2009,YangPNAS2014a,YangPNAS2014b}, and explains a variety of anomalous properties observed in heavy fermion materials \cite{YangReview2016}. But a microscopic description of the two-fluid phenomenology is still lacking, and no consensus has been reached on how exactly the $f$-electrons become delocalized \cite{LonzarichRPP2017}. 

Tremendous theoretical and experimental efforts in past decades have been focused on the so-called Kondo-destruction scenario, in which the local-itinerant transition was predicted to occur abruptly through a quantum critical point (QCP) at zero temperature \cite{Si2001,Coleman2001,Senthil2004}. While it seems to be supported experimentally by the Hall coefficient jump under magnetic field extrapolated to zero temperature in YbRh$_2$Si$_2$ \cite{YRS_Paschen2004} and the de Haas-van Alphen experiment under pressure in CeRhIn$_5$ \cite{Shishido2005}, it was lately challenged by a number of angle-resolved photoemission spectroscopy measurements showing signatures of large Fermi surfaces \cite{Kummer2015} or band hybridization above the magnetically ordered state \cite{Chen2018CeRhIn5}. In theory, the Kondo-destruction scenario could be derived under certain local or mean-field approximations, such as the dynamical large-$N$ approaches assuming independent electron baths coupled to individual impurity \cite{Komijani2019, Wang2020_local, Wang2022_fermion} and the extended dynamical mean-field theory by mapping the Kondo lattice to a single impurity Bose-Fermi Kondo model \cite{Si2001}. Since the corresponding spin-$\frac12$ single- or two-impurity problems only allow for two stable fixed points in the strong-coupling limit and the decoupling limit \cite{Sengupta2000, Cai2019, Rech2006}, these approaches unavoidably predicted a single QCP associated with Kondo destruction.

However, there is no a priori reason to assume such a local impurity mapping to be always valid for Kondo lattice systems in which all spins are spatially correlated and coupled to a common shared bath. For example, in CePdAl \cite{Zhao2019}, geometric frustration may promote quantum fluctuations of local spins so that the single QCP is replaced by an intermediate quantum critical phase at zero temperature \cite{Wang2021_nonlocal,Wang2022_Z2}. Numerically, density-matrix renormalization group (DMRG) calculations of the one-dimensional (1D) Kondo lattice have predicted an intermediate phase with neither large nor small Fermi surfaces \cite{Eidelstein2011}. For 2D Kondo lattice, both quantum Monte Carlo (QMC) simulations \cite{Assaad2021, Watanabe2007} and the dynamical cluster approach \cite{Assaad2008} have suggested continuous existence of Kondo screening inside the magnetic phase. In particular, an effective nonlocal Kondo interaction has recently been proposed using an improved Schwinger boson approach with full momentum-dependent self-energies, yielding intermediate ground states with partially enlarged electron Fermi surfaces \cite{Wang2021_nonlocal,Wang2022_Z2, Wang2022_3IK}. It is therefore necessary to go beyond the local or mean-field approximations and explore in a more rigorous manner how $f$-electrons may evolve once nonlocal interaction effects are taken into account.

In this work, we extend the concept of Kondo interaction to an extreme case where the nonlocal scattering between conduction electrons and spins has an infinite interacting range such that it becomes local in the momentum space. We further include a Heisenberg-like term in the momentum space to mimic the Kondo-RKKY competition in heavy fermion materials. Similar to the Hatsugai-Kohmoto model with a $\textbf{ k}$-space Hubbard-$U$ interaction \cite{HK1992, Phillips2020, Phillips2022, YinZhong2022, YuLi2022}, our proposed $\textbf{ k}$-space Kondo-Heisenberg model is exactly solvable. This allows us to overcome uncertainties in previous studies introduced by either analytical approximations or numerical ambiguities and extract decisive information on potential physical effects of nonlocal correlations. We find many interesting features such as spin-charge separated excitations, coexistence of Kondo singlets and spin singlets, and continuous evolution of the Fermi surfaces. Our results yield useful insight into the microscopic description of two-fluid behaviors, highlight the rich consequences of nonlocal Kondo scattering, and provide an unambiguous counterexample to the local Kondo-destruction scenario.

\section{Results}
\subsection{The $\textbf{k}$-space Kondo-Heisenberg model} 

We begin by constructing the following Hamiltonian, 
\begin{eqnarray}
H&=&\frac{1}{2}\sum_{\bf k}H_{\bf k},\notag \\
H_{\bf k}&=&(\epsilon_{\bf k}-\mu)(n_{\bf k}+n_{\bf -k})+J_K({\bf s}_{\bf k}\cdot {\bf S}_{\bf k}+{\bf s}_{\bf -k}\cdot {\bf S}_{\bf -k})\notag \\
& &+J_H{\bf S}_{\bf k}\cdot {\bf S}_{\bf -k},
\label{eq:H1}
\end{eqnarray}
where $n_{\mathbf{k}}=\sum_\alpha c_{\mathbf{k}\alpha}^\dagger c_{\mathbf{k}\alpha}$ is the electron occupation number at momentum ${\bf k}$, $\mu$ is the chemical potential, and $\epsilon_{\bf k}=\epsilon_{\bf -k}$ is the  electron dispersion relation. The electron spin $\mathbf{s}_{\bf k}=\frac{1}{2}\sum_{\alpha\beta}c_{\mathbf{k}\alpha}^\dagger \bm{\sigma}_{\alpha \beta}c_{\mathbf{k}\beta}$ and the local spin $\mathbf{S}_{\bf k}$ are both defined in the momentum space. Note that ${\bf S_k}$ is not the Fourier transform of the spin operator in the coordinate space, but should rather be viewed as that of an ``$f$-electron'' localized in the momentum space. In the pseudofermion representation, this corresponds to $\mathbf{S}_{\bf k}=\frac{1}{2}\sum_{\alpha\beta}f_{\mathbf{k}\alpha}^\dagger \bm{\sigma}_{\alpha \beta}f_{\mathbf{k}\beta}$ under the constraint $\sum_{\alpha}f_{\mathbf{k}\alpha}^\dagger f_{\mathbf{k}\alpha}=1$. It is immediately seen that the Kondo interaction is highly nonlocal by Fourier transform to the coordinate space, $\frac{J_K}{2}\sum_{iji'j'}c_{i\alpha}^\dagger c_{j\beta}f_{i'\beta}^\dagger f_{j'\alpha}\delta_{\mathbf{r}_{i}-\mathbf{r}_{j}, \mathbf{r}_{j'}-\mathbf{r}_{i'}}$. A similar form of nonlocal Kondo interaction has been suggested to emerge in the quantum critical regime  and play an important role in strongly frustrated Kondo systems \cite{Wang2021_nonlocal,Wang2022_Z2, Wang2022_3IK}.

\begin{table}[b]
\begin{tabular}{p{1.5cm}<{\centering} p{1cm}<{\centering} p{2cm}<{\centering} p{2.4cm}<{\centering} } \hline \hline
\specialrule{0em}{2pt}{2pt}
${\bf k}$    & $(n_{\bf k}, n_{\bf -k})$ & $E_{\bf k}$ & Ground State  \\
\specialrule{0em}{2pt}{2pt} 
 \hline
\specialrule{0em}{2pt}{2pt} 
$\epsilon_{\bf k}-\mu>\zeta$ &  (0,0)  & $-\frac{3}{4}J_H$     &      $\left| 0 0\right\rangle \otimes \left|\text{SS}\right\rangle$ \\
\specialrule{0em}{4pt}{4pt} 
$|\epsilon_{\bf k}-\mu|<\zeta$ & (1,1) & $2(\epsilon_{\bf k}-\mu)-\frac{J_K+\tilde{J}}{2}-\frac{J_H}{4}$   &   $a\left|\text{KS}\right\rangle_{\bf k}\otimes \left|\text{KS}\right\rangle_{\bf -k}+b\left|\text{ss}\right\rangle \otimes \left|\text{SS}\right\rangle $ \\  
\specialrule{0em}{4pt}{4pt} 
$\epsilon_{\bf k}-\mu<-\zeta$ & (2,2) & $4(\epsilon_{\bf k}-\mu)-\frac{3J_H}{4}$ &  $\left| 22\right\rangle \otimes \left|\text{SS}\right\rangle$ \\
\specialrule{0em}{2pt}{2pt}  \hline \hline
\end{tabular}
\caption{The ground states of $H_{\bf k}$. $E_{\bf k}$  is the ground state energy. $\left|00\right\rangle$ and $\left|22\right\rangle$ denote the empty and fully occupied electron states at ${\bf k}$ and ${\bf -k}$.  $\left|\text{ss}\right\rangle$ ($\left|\text{SS}\right\rangle$) denotes the spin singlet formed by the two electrons (local spins) at ${\bf k}$ and ${\bf -k}$, while $\left|\text{KS}\right\rangle_{\bf k}$ denotes the Kondo singlet at ${\bf k}$. The ratio between the coefficients $a$ and $b$ is  $2J_K/(J_H+\tilde{J}-2J_K)$. }
\label{tab1}
\end{table}

\begin{figure}[t]
\centering\includegraphics[scale=0.57]{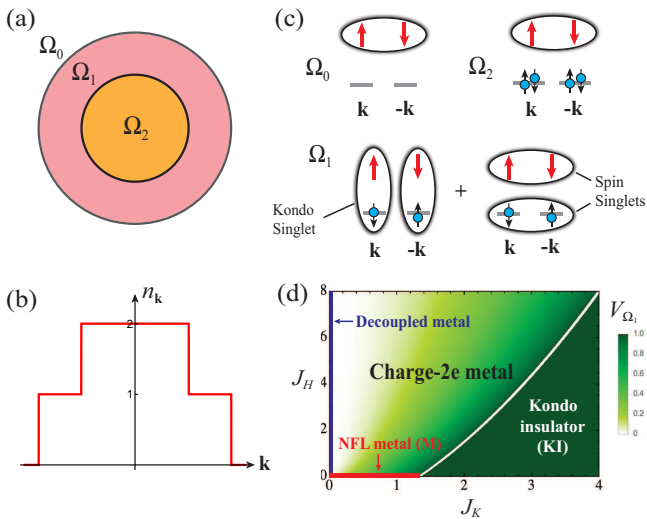}
\caption{The ground state of $\textbf{k}$-space Kondo-Heisenberg model. (\textit{A}) The momentum space contains three regions with different electron occupation number shown in (\textit{B}). (\textit{C}) Ground states of $H_{\bf k}$ in each momentum region. The red arrows and blue balls with a black arrow denote the local spins and conduction electrons, respectively. The ellipses represent  the entangled Kondo singlet or spin singlet. (\textit{D}) The ground state phase diagram at $\mu=0$, showing different phases. The color represents the volume of the singly occupied region $\Omega_1$. }
\label{fig:KH}
\end{figure}

The above model is exactly solvable, since the total Hilbert space can be divided into many small and independent subspaces by each conserved $H_{\bf k}$. The local Hilbert space at each momentum point contains 8 states constructed by 4 electron states ($\left| 0\right\rangle$, $\left|\uparrow \right\rangle$, $\left|\downarrow\right\rangle$, $\left| 2\right\rangle$) and 2  spin states ($\left| \Uparrow \right\rangle$, $\left| \Downarrow \right\rangle$), so $H_{\bf k}$ has a total number of 64 eigenstates and can be exactly diagonalized. These states are further classified into different sectors by the electron numbers $(n_{\bf k}, n_{\bf -k})$. Depending on the relative magnitudes of $\epsilon_{\bf k}-\mu$ and $\zeta\equiv(J_K-J_H+\tilde{J})/4$, where $\tilde{J}=\sqrt{J_H^2-2J_HJ_K+4J_K^2}$, we may find the ground state of $H_{\bf k}$ among three possibilities: 1) for $\epsilon_{\bf k}-\mu>\zeta$, one has $(n_{\bf k}, n_{\bf -k})=(0,0)$, and $\mathbf{S}_{\bf k}$, $\mathbf{S}_{\bf -k}$ form a spin singlet; 2) for $|\epsilon_{\bf k}-\mu|<\zeta$, $(n_{\bf k}, n_{\bf -k})=(1,1)$, and the ground state is a superposition between Kondo singlets and spin singlets, as shown in Table \ref{tab1} and Fig. \ref{fig:KH}(c); 3) for $\epsilon_{\bf k}-\mu<-\zeta$,  one has $(n_{\bf k}, n_{\bf -k})=(2,2)$, and the two $\textbf{k}$-local spins form a singlet.  Other sectors like $(n_{\bf k}, n_{\bf -k})=(0,1)$ and $(1, 2)$ only contribute to excited states (see \textit{Appendix A}). The momentum space  is therefore separated into three different regions, $\Omega_0$, $\Omega_1$ and $\Omega_2$, corresponding to $n_{\bf k}=0,1, 2$, as illustrated in Figs. \ref{fig:KH}(a) and \ref{fig:KH}(b). The ground state of $H$ is simply a direct product of the above three states at different ${\bf k}$.

Many interesting properties arise from the existence of the singly occupied region $\Omega_1$, which seems to be a general feature of models with $\textbf{ k}$-space local interactions \cite{HK1992, Baskaran1991, Baskaran1994, TaiKaiNg2020}. The  volume of $\Omega_1$, defined as $V_{\Omega_1}=\frac{1}{\mathcal{N}}\sum_{\bf k}\theta(\zeta-|\epsilon_{\bf k}-\mu|)$ where $\mathcal{N}$ is the total number of ${\bf k}$ points, is shown in Fig. \ref{fig:KH}(d), which maps out the phase diagram on the $J_H$-$J_K$ plane. For simplicity, we have assumed $\epsilon_{\bf k}=k^2/2\pi-1$, $\mu=0$, and $\epsilon_{\bf k}-\mu \in [-1,1]$. The momentum average is then $\frac{1}{\mathcal{N}}\sum_{\bf k}\equiv \int_{|{\bf k}|<k_\Lambda}d^2{\bf k}/(2\pi)^2$, where $k_\Lambda=2\sqrt{\pi}$ is the momentum cutoff corresponding to a Brillouin zone volume $(2\pi)^2$. At $J_K=0$,  one has $V_{\Omega_1}=0$, and the conduction electrons are completely decoupled from the ``${\bf k}$-space valence bond state'' formed by the local spins \cite{TaiKaiNg2020}, hence the name decoupled metal. For $J_K$ and $J_H$ satisfying $\zeta\geq 1$ (below the white curve in Fig. \ref{fig:KH}(d)), one has $V_{\Omega_1}=1$, such that all spins are Kondo screened by conduction electrons. This is the Kondo insulator (KI) phase with an insulating gap around the Fermi energy. In between, one has  $0<V_{\Omega_1}<1$,  and the system is in a charge-2$e$ metal with gapped single-particle excitations but gapless two-particle (Cooper pair) excitations. As one approaches the $J_H=0$ limit from inside the charge-2$e$ metal, the single particle gap vanishes, and the system becomes a non-Fermi liquid (NFL) metal, which we denote as M.

\subsection{Excitations} 
\begin{figure}[t]
\centering\includegraphics[scale=0.46]{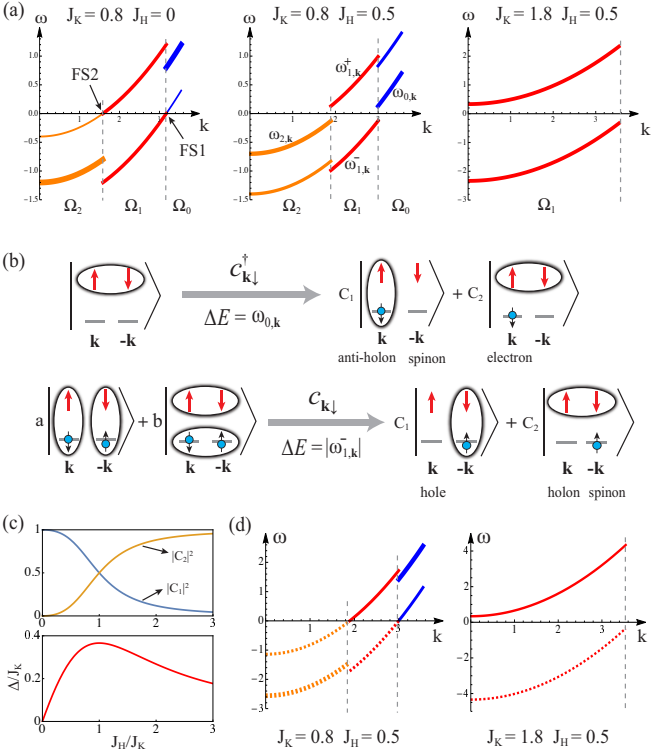}
\caption{Low-energy excitations. (\textit{A}) The poles of the single particle Green's function at $\mu=0$ for three typical values of $J_K$ and $J_H$ corresponding to the M (left), charge-2$e$ metal (middle) and KI (right) phases. The blue, red and orange curves represent the excitations with momentum  ${\bf k}\in\Omega_0$, $\Omega_1$ and $\Omega_2$, respectively, and the thickness of the curves is proportional to the spectral weight of the poles.  (\textit{B}) The physical meanings of the poles $\omega_{0,{\bf k}}$ and $\omega_{1,{\bf k}}^-$. (\textit{C}) The coefficient $C_1$ and $C_2$ (up) and the single particle gap  (bottom) as  functions of $J_H/J_K$ in the charge-2$e$ metal phase. (\textit{D}) The poles of the two-particle Green's function in the charge-2$e$ metal (left) and KI (right) phases. The dashed curves correspond to the poles with negative spectral weights. }
\label{fig:GF}
\end{figure}

The elementary excitations can be obtained exactly from the single-particle retarded Green's function defined as $G_c({\bf k},t)=-i\theta(t)\left\langle \{  c_{\mathbf{k}\alpha}(t),c_{\mathbf{k}\alpha}^\dagger \}\right\rangle$. Its explicit analytical expression at zero temperature is given in \textit{Appendix B}. The poles of the Green's function are plotted in Fig. \ref{fig:GF}(a)  in different phases, with the spectral weights represented by the thickness of the curves. Two additional poles in the $\Omega_1$ region are not shown as they have very small weights and locate far away from the Fermi energy. For $\zeta<1$,  the following poles are most close to the Fermi energy:
\begin{eqnarray}
\omega_{0,{\bf k}}&=&\epsilon_{\bf k}-\mu-\frac{J_K-2J_H+ 2\tilde{J}'}{4},\qquad   {\bf k}\in \Omega_0\notag \\
\omega_{1,{\bf k}}^{\pm}&=&\epsilon_{\bf k}-\mu\pm\frac{J_K+2\tilde{J}- 2\tilde{J}'}{4},\qquad  {\bf k}\in \Omega_1 \\
\omega_{2,{\bf k}}&=&\epsilon_{\bf k}-\mu+\frac{J_K-2J_H+ 2\tilde{J}'}{4},\qquad  {\bf k}\in \Omega_2\notag 
\label{eq:pole}
\end{eqnarray} 
where $\tilde{J}'=\sqrt{J_H^2-J_HJ_K+J_K^2}$.  Physically, $\omega_{0,{\bf k}}$ corresponds to adding one electron at ${\bf k}\in\Omega_0$, so that the system is excited from the state $\left| 0 0\right\rangle \otimes \left|\text{SS}\right\rangle$ to one of the lowest doublets of the $(n_{\bf k},n_{\bf -k})=(1,0)$ sector, for example, $C_1\left|\text{KS}\right\rangle_{\bf k}\otimes \left|\Downarrow\right\rangle_{\bf -k}+C_2 \left|\text{SS}\right\rangle\otimes \left|\downarrow\right\rangle_{\bf k}$ if the added electron has a down spin (see Fig. \ref{fig:GF}(b)). Interestingly, the component $\left|\text{KS}\right\rangle_{\bf k}\otimes \left|\Downarrow\right\rangle_{\bf -k}$ creates  a charge $-e$ excitation (anti-holon \cite{Baskaran1991}) at ${\bf k}$ and a spin-1/2 excitation (spinon) at ${\bf -k}$, while the component $\left|\text{SS}\right\rangle\otimes \left|\downarrow\right\rangle_{\bf k}$ creates an electron excitation at $\bf k$. The former indicates spin-charge separated excitations that dominate at small $J_H/J_K$ due to the vanishing weight $|C_2|^2$ in the $J_H\rightarrow 0$ limit  as shown in Fig. \ref{fig:GF}(c). Similarly, the pole $\omega_{1,{\bf k}}^{-}$ corresponds to removing one electron at  ${\bf k}\in\Omega_1$, and the resulting excited state is a superposition between a hole excitation at $\bf k$ (with coefficient $C_1$), and a holon-spinon pair located at opposite momentum points (with coefficient $C_2$).  The poles $\omega_{1,{\bf k}}^+$ and $\omega_{2,{\bf k}}$ have similar  physical meanings,  but with the empty states in Fig. \ref{fig:GF}(b) replaced by the double-occupied states.

In the charge-2$e$ metal, as shown in Fig. \ref{fig:GF}(a), the poles $\omega_{0,{\bf k}}$ and $\omega_{1,{\bf k}}^{-}$ are separated by a direct energy gap at the $\Omega_0$-$\Omega_1$ boundary, and the same for  $\omega_{1,{\bf k}}^+$ and $\omega_{2,{\bf k}}$ at the $\Omega_1$-$\Omega_2$ boundary. We find the gap follows a scaling function $\Delta/J_K=\frac{1}{2}[z+(z^2-2z+4)^{1/2}-2(z^2-z+1)^{1/2}]$, with $z=J_H/J_K$. It vanishes in the limit $J_H\rightarrow 0$, leading to two ``Fermi surfaces'' in the M phase, as denoted by FS1 and FS2 in Fig. \ref{fig:GF}(a).  However, these are not usual electron Fermi surfaces, in the sense that moving an electron from one side of the Fermi surface to the other causes spin-charge separation. Therefore, the M phase at $J_H=0$ is  actually a NFL metal. We will see that even for  $J_H>0$, the physics should be qualitatively identical to the M phase at temperatures higher than the single particle gap of the charge-2$e$ metal ground state.

Inside the KI phase, both $\Omega_0$ and $\Omega_2$  disappear, and the single particle gap becomes an indirect gap between  $\omega_{1,{\bf k}}^+$  and $\omega_{1,{\bf k}}^-$. This gap remains open in the $J_H\rightarrow 0$ limit, and has a different nature  from that of   the charge-2$e$ metal. Their difference becomes more clear when we consider the two-particle Green's function, $G_{b}(\mathbf{k},t)=-i\theta(t)\left\langle  [b_{\bf k}(t), b_{\bf k}^\dagger ] \right\rangle$, where $b_{\bf k}^\dagger=\frac{1}{\sqrt{2}}(c_{{\bf k}\uparrow}^\dagger c_{{\bf -k}\downarrow}^\dagger-c_{{\bf k}\downarrow}^\dagger c_{{\bf -k}\uparrow}^\dagger)$ creates a singlet pair of electrons (a Cooper pair) \cite{Baskaran1994}.  As shown in Fig. \ref{fig:GF}(d), $G_b({\bf k},\omega)$ is gapped in the KI phase but gapless in the charge-2$e$ metal. This means, inside the charge-2$e$ metal, adding or removing a singlet pair of electrons at ${\bf k}$ and ${\bf -k}$ costs no energy  if ${\bf k}$ locates exactly at the  $\Omega_0$-$\Omega_1$ or $\Omega_1$-$\Omega_2$ boundaries, indicating Cooper pairs rather then electrons being its elementary charge carriers. However, because our simple model does not contain scatterings between Cooper pairs, this state can only be viewed as a completely quantum disordered superconductor without long-range phase coherence \cite{TaiKaiNg2020, Kapitulnik2019}.

\subsection{Two-fluid behavior} The fact that the ground state involves a superposition of the Kondo singlets and local spin singlets in the momentum space is reminiscent of the two-fluid model of heavy fermion materials, in which an ``order parameter'' $f(T)=\min\{1, f_0(1-T/T^*)^{3/2}\}$ was found to characterize the fraction of hybridized $f$-moments over a broad temperature range, with $f_0$ reflecting the strength of collective hybridization (or collective Kondo entanglement) \cite{YangPRL2008, YangPNAS2012}. $f_0\geq 1$ indicates full screening below some characteristic temperature where $f(T)$ reaches unity, while  $0<f_0<1$ implies that  a fraction of $f$-electrons may remain unhybridized even down to zero temperature if the scaling is not interrupted by other orders. The two-fluid model captures a large amount of experimental properties of heavy fermion metals \cite{YangReview2016}, but its microscopic theory remains to be explored \cite{LonzarichRPP2017}. 

\begin{figure}[t]
\centering\includegraphics[scale=0.5]{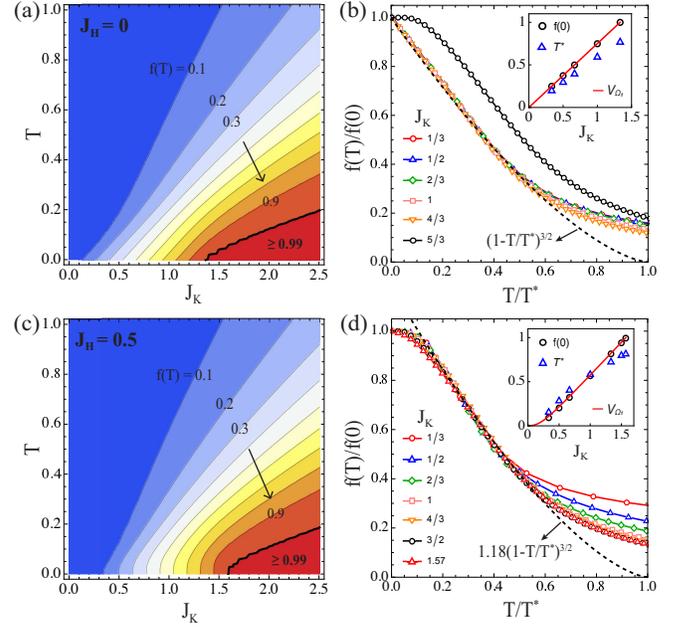}
\caption{The two-fluid ``order parameter''. (\textit{A}) A contour plot of the two-fluid ``order parameter'' $f(T)$ for $J_H=0$. (\textit{B}) $f(T)/f(0)$ as a function of $T/T^*$ for $J_H=0$ and different $J_K$. The dashed curve shows the phenomenological scaling function $(1-T/T^*)^{3/2}$ for comparison. The inset shows $f(0)$, $T^*$, and $V_{\Omega_1}$ as functions of $J_K$. (\textit{C})(\textit{D}) The same as (\textit{A})(\textit{B}), but with $J_H=0.5$. The dashed curve in (\textit{D}) corresponds to $1.18(1-T/T^*)^{3/2}$. } 
\label{fig:TF}
\end{figure}

To see how two-fluid behavior may emerge in our exactly solvable model, we introduce the projector  $P_{\bf k}=\left| \bf{K}\right\rangle \left\langle \bf{K} \right|$ with $\left| \bf{K}\right\rangle =\frac{1}{2}\left(\left|\uparrow \Downarrow\right\rangle-\left|\downarrow \Uparrow\right\rangle\right)_{\bf k}\left(\left|\uparrow \Downarrow\right\rangle-\left|\downarrow \Uparrow\right\rangle\right)_{\bf -k}$, and its momentum average $P=\frac{1}{\mathcal{N}}\sum_{\bf k}P_{\bf k}$. This gives a two-fluid ``order parameter'',
\begin{eqnarray}
f(T)=\frac{\text{Tr}[e^{-H/T}P]}{\text{Tr}[e^{-H/T}]}=\frac{1}{\mathcal{N}}\sum_{\bf k}\frac{\text{Tr}[e^{-H_{\bf k}/T}P_{\bf k}]}{\text{Tr}[e^{-H_{\bf k}/T}]},
\label{eq:TF}
\end{eqnarray}
which reflects the fraction of Kondo singlet formation in the momentum space. With this definition, it is easy to show that a physical observable can in principle also be divided  into a two-fluid form $\langle O\rangle=f\langle O\rangle_{P}+(1-f)\langle O\rangle_{1-P}$.

Figures \ref{fig:TF}(a) and \ref{fig:TF}(c) show the contour plots of the calculated $f(T)$ at $J_H=0$ and $0.5$, respectively. In general, we see $f(T)$ increases with decreasing temperature and saturates to a finite zero temperature value $f(0)$. For $J_H=0$, $f(0)$ increases linearly from 0 to 1 with increasing $J_K$,  and stays at unity for $J_K>4/3$ (inside the KI phase). For $J_K<4/3$ (inside the M phase), $f(T)$ follows a universal scaling function $f(T)/f(0)=F(T/T^*)$, as shown in Fig. \ref{fig:TF}(b). Quite remarkably, the low temperature part of $F(T/T^*)$ can be well approximated by the function $(1-T/T^*)^{3/2}$. At high temperatures, its smooth evolution reflects a crossover rather than a phase transition of the delocalization with temperature. For $J_K>4/3$, $f(T)$ grows to unity already at a finite temperature, in good agreement with the expectation of the two-fluid picture \cite{YangPNAS2012}. The results for $J_H=0.5$ are slightly different. We find for small $J_H$, $f(T)$ already stays constant below certain temperature before it reaches unity. This is due to the energy gap of the charge-2$e$ metal that interrupts the two-fluid scaling. Above the gap, $f(T)$ follows the same two-fluid scaling behavior over a broad intermediate temperature range, as shown in Fig. \ref{fig:TF}(d). The similar two-fluid behavior clearly indicates that the intermediate temperature physics above the charge-2$e$ metal is controlled by the NFL M phase with partial Kondo screening rather than the charge-2$e$ metal. This may have important implications for real materials, where the scaling is often interrupted or even suppressed ($f$-electron relocalization) by magnetic, superconducting, or other long-range orders. A second observation is that $f(0)$ as a function of $J_K$ is nearly identical to the volume of single-occupied region, as shown by the red line in the inset of Figs. \ref{fig:TF}(b) and \ref{fig:TF}(d). This confirms the previous speculation of an intimate relation between the two-fluid ``order  parameter'' and the partial Kondo screening at zero temperature \cite{YangPNAS2012}. The quantum state superposition revealed in the exactly solvable model may also be the microscopic origin of the two-fluid phenomenology widely observed in real heavy fermion materials.

\subsection{Luttinger's theorem} The Luttinger's theorem provides an important criterion for Landau's Fermi liquid description of interacting electron systems \cite{LuttingerWard1960, Luttinger1960,Oshikawa2000}. It states that the volume enclosed by the Fermi surface should be equal to the number of conduction electrons per unit cell. Mathematically, it is often quoted as \cite{Phillips2020,Dzyaloshinskii2003, Phillips2007,Dave2013}
\begin{equation}
V_{\text{LC}}\equiv \frac{2}{\mathcal{N}}\sum_{\bf k}\theta(\text{Re}G_c({\bf k},0))=n_c, 
\label{eq:LC}
\end{equation}
where the factor 2 arises from the up and down spins, and $n_c=\frac{1}{\mathcal{N}}\sum_{\bf k} \langle n_{\bf k}\rangle$ is the electron density. For a Fermi liquid metal, $\text{Re}G_c({\bf k},0)$ changes its sign only at the Fermi surface by passing through infinity, and hence Eq. (\ref{eq:LC}) reduces to the simple Fermi volume statement. It was later suggested that Eq. (\ref{eq:LC}) can also be applied  to systems without quasiparticle poles \cite{Dzyaloshinskii2003, Tsvelik2006}, such as the Mott insulator. In that case, $\text{Re}G_c({\bf k},0)$ changes sign by passing through its zeros, which form a Luttinger surface \cite{Dzyaloshinskii2003}.  However, the Luttinger surface of a Mott insulator was found to depend on the arbitrary choice of $\mu$, such that Eq. (\ref{eq:LC}) only holds with the presence of particle-hole symmetry \cite{Phillips2007, Rosch2007}. This suggests a failure of Eq. (\ref{eq:LC}) and possibly nonexistence of the Luttinger-Ward functional in these strongly correlated systems \cite{Phillips2007,Dave2013,Kozik2015}.

\begin{figure}[t]
\centering\includegraphics[scale=0.47]{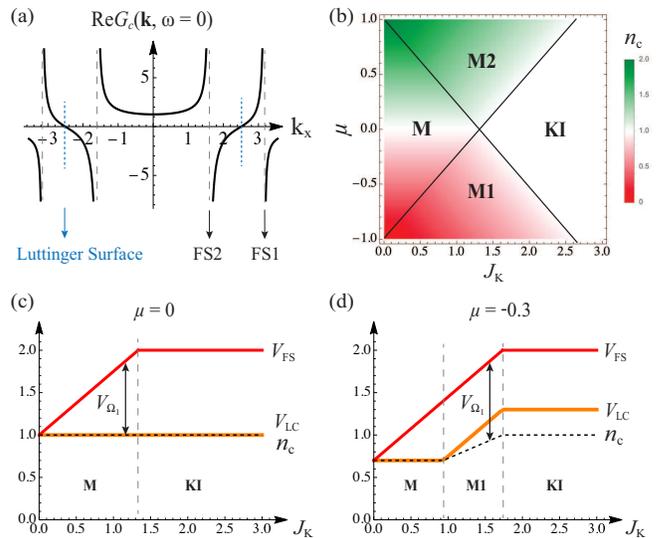}
\caption{The Fermi volume evolution and the Luttinger's theorem. (\textit{A}) Real part of the Green's function $G_c({\bf k},0)$ at $\mu=0$, $J_H=0$, $J_K=0.8$, showing both the Luttinger surface and the Fermi surfaces (FS1 and FS2). (\textit{B}) The electron density as a  function of $\mu$ and $J_K$ at $J_H=0$.  M, M1 and  M2 denote different metallic phases, and KI is the Kondo insulator phase. (\textit{C})(\textit{D}) Evolution of $V_{\text{FS}}$, $V_{\text{LC}}$, and $n_c$ with increasing $J_K$ at $\mu=0$ and -0.3.  }
\label{fig:LT}
\end{figure}

Here, we demonstrate based on our model that the naive Fermi volume counting is in fact better than the Luttinger count $V_{\text{LC}}$ on representing the electron density. As shown in Fig. \ref{fig:LT}(a), the real part of the Green's function $\text{Re}G_c({\bf k},0)$ at $J_H=0$  reveals a Luttinger surface inside $\Omega_1$ and two Fermi surfaces at the boundaries of $\Omega_1$ and $\Omega_2$. Therefore, we can define the Fermi volume as  $V_{\text{FS}}\equiv 2(V_{\Omega_1}+V_{\Omega_2})$, and study its relation to the electron density. To do this, we first calculate $n_c$ as a function of $J_K$ and $\mu$ at $J_H=0$. The result is shown in Fig. \ref{fig:LT}(b).  For nonzero $\mu$,  there exist another two metallic phases, M1 and M2, where one of the two Fermi surfaces disappears due to the absence of $\Omega_2$ or $\Omega_0$ region. Both M1 and M2 will open a single particle gap by turning on a finite $J_H$, and become another two charge-2$e$ metals. These phases have  qualitatively the same physical properties with their counterparts at $\mu=0$, and hence will not be discussed in detail.

In Figs. \ref{fig:LT}(c) and  \ref{fig:LT}(d), we compare $V_{\text{LC}}$ and  $V_{\text{FS}}$ with the electron density $n_c$ as functions of $J_K$ at $\mu=0$ and $\mu=-0.3$, respectively. At $\mu=0$, we found $V_{\text{LC}}=n_c=1$ for both the M and KI phases.  On the other hand, $V_{\text{FS}}$ evolves continuously from  $n_c$  at $J_K=0$ to $n_c+1$  in the KI phase. The deviation $V_{\text{FS}}-n_c$ is exactly equal to the volume of $\Omega_1$. In fact, the identity 
\begin{equation}
V_{\text{FS}}\equiv 2(V_{\Omega_1}+V_{\Omega_2})=n_c+V_{\Omega_1} \label{eq:VFS}
\end{equation}
holds for arbitrary $\mu $ and $J_K$, since the electron density can always be written as $n_c=V_{\Omega_1}+2V_{\Omega_2}$. Equation (\ref{eq:VFS}) correctly accounts for the Fermi surface enlargement due to the  Kondo screening effect, an important feature of the Kondo lattice \cite{Oshikawa2000}. By contrast, the deviation $V_\text{LC}-n_c$ depends explicitly on the chemical potential in the M1, M2, and KI phases,  as shown in Fig. \ref{fig:LT}(d) for $\mu=-0.3$. The parabolic free electron dispersion leads to $V_{\text{LC}}=n_c$ in the M phase for all $\mu$, which is generally  not true for other forms of $\epsilon_{\bf k}$.  In fact, one can derive analytically (see \textit{Appendix C})
\begin{equation}
V_{\text{LC}}=n_c+\frac{1}{\mathcal{N}}\sum_{{\bf k}\in\Omega_1}\text{sgn}(\epsilon_{\bf k}-\mu),
\end{equation}
which points to a general violation of Eq. (\ref{eq:LC}) when $\Omega_1$ is present.  However, this equation does not reflect the Fermi surface enlargement due to the Kondo screening effect, and is not as useful as Eq. (\ref{eq:VFS}) due to its explicit dependence on  $\epsilon_{\bf k}$ and $\mu$.

It should be noted that Eq. (\ref{eq:VFS}) has the same form as the generalized Luttinger sum rule derived in the Schwinger boson formalism of the Kondo lattice, where  $V_{\Omega_1}$ corresponds to the volume of an emergent holon Fermi surface \cite{Wang2021_nonlocal, Wang2022_Z2, coleman2005sum}. In both cases,  an intermediate phase with $0<V_{\Omega_1}<1$ is allowed, featured with partial (nonlocal) Kondo screening of local spins and gapless spinon and holon excitations, which is completely different from the  Kondo-destruction scenario where $V_{\text{FS}}$ jumps from $n_c$ to $n_c+1$ through a local QCP. This partial screening in the momentum space should be distinguished from those studied in the coordinate space \cite{Motome2010PKS}, which is always accompanied by broken translational symmetry.

%%%%%%%%%%%%%%%%%%%%%%%%%%%%%%%%%%%%%%%%%%%%%%%%%%%%%%%%%%%%%%
\section{Discussion}
We briefly discuss to what extent our toy model reflects the true physics of correlated $f$-electron systems. First, the momentum space local spins can be originated from an infinitely large Hatsugai-Kohmoto (HK) interaction between $f$-electrons, $U\sum_{\bf k}n_{{\bf k}\uparrow}^f n_{{\bf k}\downarrow}^f$. Although being a simplification of the Hubbard model, the HK model has recently been shown to capture the essential physics of Mottness and some important high-$T_c$ features upon doping \cite{Phillips2020,Phillips2022}. As suggested in Ref. \cite{Phillips2022}, this is possibly because the HK interaction is the most relevant part of the Hubbard interaction that drives the system away from the Fermi liquid fixed point to the Mott insulator. In fact, a perfect  single-occupancy constraint on every lattice site ($n_{i}^f=1$) must also imply the single-occupancy at each momentum point ($n_{\bf k}^f=1$). Therefore, we believe our model does capture the essential physics of strongly correlated $f$-electrons. Second, the Kondo term of our model contains a particular form of nonlocal Kondo interaction proposed in recent Schwinger boson theories of Kondo lattices with strong quantum fluctuation or geometric frustration \cite{Wang2021_nonlocal,Wang2022_Z2}, $J_K(|{\bf r}_i-{\bf r}_j|)c_{i\alpha}^\dagger c_{j\beta} f_{j\beta }^\dagger f_{i\alpha}$. It is related to the term $c_{i\alpha}^\dagger \bm{\sigma}_{\alpha\beta}c_{j\beta}\cdot \bm{S}_i\times \bm{S}_j$ that emerges naturally upon renormalization group from a Kondo lattice, and may become important in the quantum critical region \cite{Wang2022_3IK}. 

In summary, we have constructed an exactly solvable Kondo-Heisenberg model in momentum space. This model displays many interesting properties: 1) it realizes a charge-2$e$ metal phase with gapped single particle excitations but gapless Cooper pair excitations; 2)  as the Heisenberg interaction vanishes, the charge-2$e$ metal becomes a NFL metal featured with a partially enlarged Fermi volume; 3) both the charge-2$e$ metal and the NFL metal show universal two-fluid behaviors at finite temperatures, reflecting partial Kondo screening of local spins. All these interesting properties arise from the highly nonlocal Kondo interaction in real space, which might play an important role in heavy fermion systems. Our results may help to understand the experimentally observed NFL quantum critical phase in CePdAl \cite{Zhao2019}.  For other materials like YbRh$_2$Si$_2$, such nonlocal physics might become important in the quantum critical region, causing the smooth evolution of the Fermi surface.

\section*{ACKNOWLEDGMENT}

This work was supported by the National Natural Science Foundation of China (Grants No. 12174429, No. 11974397),  the National Key R\&D Program of China (Grant No. 2022YFA1402203), and the Strategic Priority Research Program of the Chinese Academy of Sciences (Grant No. XDB33010100).

\appendix

%\section{Materials and Methods}

\section{Exact diagonalization}\label{sec:ED}
The 64-dimensional Hilbert space of $H_{\bf k}$ can be divided into 9 subspaces according to the electron number $n_{\bf k}$ and $n_{\bf -k}$, 
\begin{eqnarray}
(n_{\bf k}, n_{\bf -k})&=&(0,0), (2,0), (0,2), (2,2) \qquad d=4 \notag \\
(n_{\bf k}, n_{\bf -k})&=&(1,0), (0,1), (1,2), (2,1) \qquad d=8 \notag \\
(n_{\bf k}, n_{\bf -k})&=&(1,1) \qquad d=16
\end{eqnarray}
where $d$ is the dimension of each subspace. To diagonalize the subspaces, we use the basis $\left|\phi_{\bf k}\phi_{\bf -k}S_{\bf k}^z S_{\bf -k}^z \right\rangle$ to compute the matrix elements, where $\phi_{\bf k}=0,\uparrow, \downarrow, 2$ denotes the four  electron states and $S_{\bf k}^z=\Uparrow, \Downarrow$ denotes the local spin states. The lowest eigenstates  within each subspace are listed in Table \ref{tab2}.   By comparing the lowest eigenenergy $E_{n_{\bf k}, n_{\bf -k}}$ of different subspaces, one obtains the ground states of $H_{\bf k}$ listed in Table \ref{tab1}.

\begin{table*}[t]
\caption{The eigenstates of $H_{\bf k}$ with the lowest energy $E_{n_{\bf k}, n_{\bf -k}}$ in each subspace labelled by $(n_{\bf k}, n_{\bf -k})$. For simplicity, the states are not normalized. The eigenstates for the $(2,0), (0,1), (2,1)$ subspaces can be obtained from those of $(0,2), (1,0), (1,2)$ by a symmetry transformation ${\bf k}\leftrightarrow {\bf -k}$.  We have defined  $\tilde{J}=\sqrt{J_H^2-2J_HJ_K+4J_K^2}$ and $\tilde{J}'=\sqrt{J_H^2-J_HJ_K+J_K^2}$.}
\begin{tabular}{p{1.4cm}<{\centering} p{4cm}<{\centering} p{11cm}<{\centering} } \hline \hline
\specialrule{0em}{2pt}{2pt}
 $(n_{\bf k}, n_{\bf -k})$ & $E_{n_{\bf k}, n_{\bf -k}}$ &  Eigenstate  \\
\specialrule{0em}{1pt}{1pt} 
 \hline
\specialrule{0em}{2pt}{2pt} 
 (0,0)  & $-\frac{3}{4}J_H$     &      $\left|00\Uparrow \Downarrow\right\rangle-\left|00\Downarrow \Uparrow\right\rangle$ \\
  \specialrule{0em}{2pt}{2pt}
 (0,2) & $2(\epsilon_{\bf k}-\mu)-\frac{3}{4}J_H$ &  $\left|02\Uparrow \Downarrow\right\rangle-\left|02\Downarrow \Uparrow\right\rangle$ \\
 \specialrule{0em}{2pt}{2pt} 
 (2,2) & $4(\epsilon_{\bf k}-\mu)-\frac{3}{4}J_H$ &  $\left|22\Uparrow \Downarrow\right\rangle-\left|22\Downarrow \Uparrow\right\rangle$ \\
\specialrule{0em}{2pt}{2pt} 
 (1,0) & $\epsilon_{\bf k}-\mu-\frac{J_H+J_K+2\tilde{J}'}{4}$ &  $(J_K+\tilde{J}'-J_H)(\left|\uparrow 0\Downarrow \Downarrow\right\rangle-\left|\downarrow 0\Uparrow \Downarrow\right\rangle )-J_H(\left|\downarrow 0\Uparrow \Downarrow\right\rangle-\left|\downarrow 0\Downarrow \Uparrow\right\rangle)$ \\
 \specialrule{0em}{2pt}{2pt} 
  &  &  $(J_H+\tilde{J}'-J_K)(\left|\uparrow 0\Uparrow \Downarrow\right\rangle-\left|\uparrow 0\Downarrow \Uparrow\right\rangle )-J_K(\left|\uparrow 0\Downarrow \Uparrow\right\rangle-\left|\downarrow 0\Uparrow \Uparrow\right\rangle)$ \\
  \specialrule{0em}{2pt}{2pt} 
 (1,2) & $3(\epsilon_{\bf k}-\mu)-\frac{J_H+J_K+2\tilde{J}'}{4}$ &  $(J_K+\tilde{J}'-J_H)(\left|\uparrow 2\Downarrow \Downarrow\right\rangle-\left|\downarrow 2\Uparrow \Downarrow\right\rangle )-J_H(\left|\downarrow 2\Uparrow \Downarrow\right\rangle-\left|\downarrow 2\Downarrow \Uparrow\right\rangle)$ \\
 \specialrule{0em}{2pt}{2pt} 
 &  &  $(J_H+\tilde{J}'-J_K)(\left|\uparrow 2\Uparrow \Downarrow\right\rangle-\left|\uparrow 2\Downarrow \Uparrow\right\rangle )-J_K(\left|\uparrow 2\Downarrow \Uparrow\right\rangle-\left|\downarrow 2\Uparrow \Uparrow\right\rangle)$ \\
\specialrule{0em}{2pt}{2pt} 
 (1,1) & $2(\epsilon_{\bf k}-\mu)-\frac{J_K+\tilde{J}}{2}-\frac{J_H}{4}$   &   $2J_K(\left|\uparrow\Downarrow\right\rangle-\left|\downarrow\Uparrow\right\rangle)_{\bf k}(\left|\uparrow\Downarrow\right\rangle-\left|\downarrow\Uparrow\right\rangle)_{\bf -k}+(J_H+\tilde{J}-2J_K)(\left|\uparrow\downarrow\right\rangle-\left|\downarrow\uparrow\right\rangle)( \left|\Uparrow\Downarrow\right\rangle-\left|\Downarrow\Uparrow\right\rangle )$ \\  
\specialrule{0em}{2pt}{2pt}  \hline \hline
\end{tabular}
\label{tab2}
\end{table*}

\section{Green's function}\label{sec:GF}

The retarded single-electron Green's function can be directly calculated from its definition, leading to
\begin{equation}
G_c({\bf k},\omega)=\sum_n \frac{|\left\langle n \right | c_{{\bf k},\alpha}^\dagger \left|0\right\rangle|^2}{\omega-E_n+E_0}+\sum_n \frac{|\left\langle n \right| c_{{\bf k},\alpha} \left|0\right\rangle|^2}{\omega+E_n-E_0},
\end{equation}
where $\omega$ represents $\omega +i0^+$, and $\left| n\right\rangle$ is the $n$-th eigenstate of $H_{\bf k}$ with energy $E_n$. The explicit analytical results are
\begin{eqnarray}
G_c({\bf k}\in \Omega_0,\omega)&=&\frac{(2\tilde{J}'+2J_H-J_K)/4\tilde{J}'}{\omega -\epsilon_{\bf k}+\mu+\frac{J_K-2J_H+2\tilde{J}'}{4}} \notag \\
&&+\frac{(2\tilde{J}'-2J_H+J_K)/4\tilde{J}'}{\omega-\epsilon_{\bf k}+\mu+\frac{J_K-2J_H-2\tilde{J}'}{4}},  
\label{eq:GF1-a}
\end{eqnarray}
\begin{eqnarray}
G_c({\bf k}\in \Omega_2,\omega)&=&\frac{(2\tilde{J}'+2J_H-J_K)/4\tilde{J}'}{\omega-\epsilon_{\bf k}+\mu-\frac{J_K-2J_H+2\tilde{J}'}{4}}\notag \\
&&+\frac{(2\tilde{J}'-2J_H+J_K)/4\tilde{J}'}{\omega-\epsilon_{\bf k}+\mu-\frac{J_K-2J_H-2\tilde{J}'}{4}},
\label{eq:GF1-b}
\end{eqnarray}
\begin{eqnarray}
G_c({\bf k}\in \Omega_1,\omega)&=&\frac{[(\tilde{J}+\tilde{J}')^2-J_K^2]/8\tilde{J}\tilde{J}'}{\omega-\epsilon_{\bf k}+\mu-\frac{J_K+2\tilde{J}-2\tilde{J}'}{4}} \notag \\
& &+ \frac{[(\tilde{J}+\tilde{J}')^2-J_K^2]/8\tilde{J}\tilde{J}'}{\omega-\epsilon_{\bf k}+\mu+\frac{J_K+2\tilde{J}-2\tilde{J}'}{4}} \notag \\
& &+\frac{[J_K^2-(\tilde{J}-\tilde{J}')^2]/8\tilde{J}\tilde{J}'}{\omega-\epsilon_{\bf k}+\mu-\frac{J_K+2\tilde{J}+2\tilde{J}'}{4}}\notag\\
& &+ \frac{[J_K^2-(\tilde{J}-\tilde{J}')^2]/8\tilde{J}\tilde{J}'}{\omega-\epsilon_{\bf k}+\mu+\frac{J_K+2\tilde{J}+2\tilde{J}'}{4}}.  
\label{eq:GF1-c}
\end{eqnarray}

For the two-particle Green's function, we have 
\begin{equation}
G_b({\bf k},\omega)=\sum_n \frac{|\left\langle n \right | b_{{\bf k}}^\dagger \left|0\right\rangle|^2}{\omega-E_n+E_0}-\sum_n \frac{|\left\langle n \right| b_{{\bf k}} \left|0\right\rangle|^2}{\omega+E_n-E_0},
\end{equation}
where $b_{\bf k}^\dagger=\frac{1}{\sqrt{2}}(c_{{\bf k}\uparrow}^\dagger c_{{\bf -k}\downarrow}^\dagger-c_{{\bf k}\downarrow}^\dagger c_{{\bf -k}\uparrow}^\dagger)$ is the Cooper pair creation operator. The analytical results are
\begin{eqnarray}
G_b({\bf k}\in \Omega_0,\omega)&=&\frac{(\tilde{J}+J_H-J_K)/2\tilde{J}}{\omega-2(\epsilon_{\bf k}-\mu)+\frac{J_K-J_H+\tilde{J}}{2}}\notag \\
& &+\frac{(\tilde{J}-J_H+J_K)/2\tilde{J}}{\omega-2(\epsilon_{\bf k}-\mu)+\frac{J_K-J_H-\tilde{J}}{2}}, 
\end{eqnarray}
\begin{eqnarray}
G_b({\bf k}\in \Omega_2,\omega)&=&-\frac{(\tilde{J}+J_H-J_K)/2\tilde{J}}{\omega-2(\epsilon_{\bf k}-\mu)-\frac{J_K-J_H+\tilde{J}}{2}}\notag \\
& &-\frac{(\tilde{J}-J_H+J_K)/2\tilde{J}}{\omega-2(\epsilon_{\bf k}-\mu)-\frac{J_K-J_H-\tilde{J}}{2}},
\end{eqnarray}
\begin{eqnarray}
G_b({\bf k}\in \Omega_1,\omega)&=&\frac{(\tilde{J}+J_H-J_K)/2\tilde{J}}{\omega-2(\epsilon_{\bf k}-\mu)-\frac{J_K-J_H+\tilde{J}}{2}}\notag \\
& &- \frac{(\tilde{J}+J_H-J_K)/2\tilde{J}}{\omega-2(\epsilon_{\bf k}-\mu)+\frac{J_K-J_H+\tilde{J}}{2}}.
\end{eqnarray}

\section{Luttinger's theorem}\label{sec:LC}

In the limit $J_H=0$, the Green's functions (\ref{eq:GF1-a})-(\ref{eq:GF1-c}) reduce to
\begin{eqnarray}
G_c(\mathbf{k},\omega)^{-1}&=&\begin{cases}
\omega-\epsilon_{\bf k}+\mu-\frac{3J_K^2/16}{\omega -(\epsilon_{\bf k}-\mu-J_K/2)}, & {\bf k}\in\Omega_0 \\
\omega-\epsilon_{\bf k}+\mu-\frac{3J_K^2/16}{\omega -(\epsilon_{\bf k}-\mu+J_K/2)}, & {\bf k}\in\Omega_2 \\
\omega-\epsilon_{\bf k}+\mu-\frac{9J_K^2/16}{\omega -(\epsilon_{\bf k}-\mu)}, & {\bf k}\in\Omega_1 
\end{cases} \notag \\
&=&\omega-\epsilon_{\bf k}+\mu-\Sigma_c({\bf k},\omega).  \label{eq:Gc} 
\end{eqnarray}
The electron density is related to the time-ordered Green's function via 
\begin{equation}
n_c=\frac{2}{\mathcal{N}}\sum_{\bf k}\int_{-\infty}^{\infty}\frac{d \omega}{2\pi}G_c({\bf k}, i\omega)e^{i\omega 0^+} \label{eq:nc}
\end{equation}
where we have performed a wick rotation $\omega+i0^+ \rightarrow i\omega$ from Eq. (\ref{eq:Gc}) to obtain the time-ordered Green's function. In proving the Luttinger's theorem, one uses the following identity,
\begin{eqnarray}
G_c({\bf k},i\omega)&=&\frac{\partial}{\partial i\omega}\ln G_c({\bf k},i\omega)^{-1} \notag \\
& &+G_c({\bf k},i\omega)\frac{\partial}{\partial i\omega}\Sigma_c({\bf k},i\omega),  
\label{eq:ID}
\end{eqnarray}
which directly follows from the Dyson's equation  (\ref{eq:Gc}). Substituting the first term of the right-hand-side of Eq. (\ref{eq:ID}) into Eq. (\ref{eq:nc}) gives exactly the Luttinger's theorem Eq. (\ref{eq:LC}). Therefore Eq. (\ref{eq:LC}) is satisfied if and only  if the following integral,
\begin{eqnarray}
I_2&\equiv& \frac{2}{\mathcal{N}}\sum_{\bf k}\int_{-\infty}^{\infty}\frac{d\omega}{2\pi }G_c({\bf k},i\omega)\frac{\partial}{\partial i\omega}\Sigma_c({\bf k},i\omega) \notag \\
&=&n_c- V_\text{LC},
\end{eqnarray}
vanishes,  which was proved by Luttinger and Ward to be true to all orders of perturbation theory \cite{LuttingerWard1960}. However, in our case, from Eq. (\ref{eq:Gc}) and the following identity,
\begin{eqnarray}
\int_{-\infty}^{\infty}\frac{d\omega}{2\pi}\frac{1}{i\omega-A}\frac{1}{(i\omega-B)^2}=\frac{\text{sgn}(B)-\text{sgn}(A)}{2(A-B)^2},
\end{eqnarray}
one can derive $I_2=-\frac{1}{\mathcal{N}}\sum_{{\bf k}\in \Omega_1}\text{sgn}(\epsilon_{\bf k}-\mu)$, which is generally nonzero.  This may originate from the nonexistence of the Luttinger-Ward functional for our system, similar to the cases studied in Refs. \cite{Phillips2007,Dave2013}. 

In fact, for any strictly monotonically increasing  function $\epsilon_{\bf k}=\epsilon(k)$ within the range $k\in [0,2\sqrt{\pi}]$, one has
\begin{eqnarray}
I_2&=&\frac{1}{2\pi}\int_{\text{max}[\epsilon(0),\mu-\frac{3J_K}{4}]}^{\mu} \frac{\epsilon^{-1}(x)}{\epsilon'(\epsilon^{-1}(x))}dx \notag \\
&&-\frac{1}{2\pi}\int_{\mu}^{\text{min}[\epsilon(2\sqrt{\pi}),\mu+\frac{3J_K}{4}]}\frac{\epsilon^{-1}(x)}{\epsilon'(\epsilon^{-1}(x))}dx,
\end{eqnarray}
where $\epsilon'(x)$ and $\epsilon^{-1}(x)$ are the derivative and inverse of the function $\epsilon(x)$, respectively. For a parabolic dispersion function $\epsilon(x)=ax^2+b$, one has $\epsilon^{-1}(x)=\sqrt{(x-b)/a}$ and $\epsilon'(x)=2ax$, so that 
\begin{eqnarray}
I_2=\begin{cases}
\frac{1}{2\pi}\left(\int_{\mu-\frac{3J_K}{4}}^\mu-\int_{\mu}^{\mu+\frac{3J_K}{4}}\right)\frac{1}{2a}dx=0, & \text{M} \\
\frac{1}{2\pi}\left(\int_{b}^\mu-\int_{\mu}^{4\pi a +b}\right)\frac{1}{2a}dx=\frac{\mu-b}{2\pi a}-1, & \text{KI} 
\end{cases}
\end{eqnarray}
consistent with our numerical results for $a=1/(2\pi) $ and $b=-1$.

\end{document}